\documentclass[journal,twoside,web]{ieeecolor}
\usepackage{generic}
\usepackage{cite}
\usepackage{amsmath,amssymb,amsfonts}
\usepackage{algorithmic}
\usepackage{graphicx}
\usepackage{algorithm,algorithmic}
\usepackage{hyperref}
\hypersetup{hidelinks=true}
\usepackage{textcomp}
\usepackage{amsmath,amssymb,amsfonts,mathtools, bm}

\usepackage{graphicx}
\usepackage[font=footnotesize]{subcaption} 
\usepackage{subcaption}

\usepackage{algorithm}      
\usepackage{algorithmic}    
\usepackage{booktabs}
\usepackage{pifont}
\newcommand{\cmark}{\ding{51}} 
\newcommand{\xmark}{\ding{55}} 

\usepackage{booktabs,multirow}
\usepackage{siunitx}    
\newlength{\rocimgheight}
\newcommand{\bestnum}[1]{\textbf{#1}}
\usepackage{algorithmic}
\usepackage{cite}
\usepackage{textcomp}

\def\BibTeX{{\rm B\kern-.05em{\sc i\kern-.025em b}\kern-.08em
    T\kern-.1667em\lower.7ex\hbox{E}\kern-.125emX}}
\author{Jintao Huang}
\begin{document}
\title{Hear the Heartbeat in Phases: Physiologically Grounded Phase-Aware ECG Biometrics}
\author{Jintao Huang,
Lu Leng,~\IEEEmembership{Member, IEEE},
Yi Zhang,~\IEEEmembership{Senior Member, IEEE},
and Ziyuan Yang,~\IEEEmembership{Member, IEEE}
\thanks{\textit{This work has been submitted to the IEEE for possible publication.
Copyright may be transferred without notice, after which this version may
no longer be accessible.}}
\thanks{This work was supported in part by the National Natural Science Foundation of China under Grants 62466038; in part by the Jiangxi Provincial Key Laboratory of Image Processing and Pattern Recognition under Grants 2024SSY03111 and ET202404437; and in part by the High Performance Computing Service of Information Center at Nanchang Hangkong University. 	\textit{(Corresponding author: Lu Leng and Ziyuan Yang})}
\thanks{Jintao Huang is with the Jiangxi Provincial Key Laboratory of Image Processing and Pattern Recognition, Nanchang Hangkong University, 330063,
China (e-mail: 2520083500018@stu.nchu.edu.cn).}
\thanks{Lu Leng is with the Jiangxi Provincial Key Laboratory of Image Processing and Pattern Recognition, Nanchang Hangkong University, 330063, China (email: leng@nchu.edu.cn).}
\thanks{Yi Zhang and Ziyuan Yang are with the School of Cyber Science and Engineering, Sichuan University, Chengdu 610065, China (yzhang@scu.edu.cn and cziyuanyang@gmail.com).}
}
\maketitle

\begin{abstract}
Electrocardiography (ECG) is adopted for identity authentication in wearable devices due to its individual-specific characteristics and inherent liveness.
However, existing methods often treat heartbeats as homogeneous signals, overlooking the phase-specific characteristics within the cardiac cycle. To address this, we propose a Hierarchical Phase-Aware Fusion~(HPAF) framework that explicitly avoids cross-feature entanglement through a three-stage design.
In the first stage, Intra-Phase Representation (IPR) independently extracts representations for each cardiac phase, ensuring that phase-specific morphological and variation cues are preserved without interference from other phases.
In the second stage, Phase-Grouped Hierarchical Fusion (PGHF) aggregates physiologically related phases in a structured manner, enabling reliable integration of complementary phase information.
In the final stage, Global Representation Fusion (GRF) further combines the grouped representations and adaptively balances their contributions to produce a unified and discriminative identity representation. Moreover, considering ECG signals are continuously acquired, multiple heartbeats can be collected for each individual. We propose a Heartbeat-Aware Multi-prototype (HAM) enrollment strategy, which constructs a multi-prototype gallery template set to reduce the impact of heartbeat-specific noise and variability.
Extensive experiments on three public datasets demonstrate that HPAF achieves state-of-the-art results in the comparison with other methods under both closed and open-set settings.
\end{abstract}


\section{Introduction}
\label{sec:introduction}

Electrocardiography (ECG) has long played a fundamental role in clinical diagnostics and, with the rise of wearable devices, is increasingly recognized as a physiologically grounded biometric signal. Compared to other modalities such as face or fingerprint, ECG acquisition originates directly from cardiac activity and thus inherently provides liveness detection~\cite{1,2,3}. These characteristics make ECG a promising modality for privacy-preserving and continuous authentication~\cite{6,7,7563392, Ma2015JBHI_AFD}.



Recently, a growing body of research has investigated the use of ECG for identity recognition. These methods have progressed from handcrafted feature-based approaches to deep learning-based ones~\cite{8,9}. Most existing approaches use a single model to encode different cardiac phases. For instance, Ibtehaz \emph{\textit{et al.}}~\cite{EDITH} employed a multi-resolution Siamese network on whole beats for ECG authentication. Chu \emph{\textit{et al.}}~ \cite{PMS_RESNET} utilized a parallel multi-scale ResNet with center and margin losses to extract discriminative features. Wang \emph{\textit{et al.}}~\cite{SSL_CNN} proposed a self-supervised convolutional neural network~(CNN) framework that compares different ECG segmentation strategies. Allam \emph{\textit{et al.}}~\cite{BAED} adopted a parallel CNN-LSTM architecture to learn representations from R-peak-aligned beats. Although promising in ideal settings, these methods uniformly treat heartbeats containing the P wave, QRS complex, and ST/T(U) segments as a homogeneous sequence processed by a shared encoder.
These methods rely on the basic assumption that different phases follow the homogeneous signal patterns, ignoring their inherent structural and physiological heterogeneity. This leads to poor discriminative feature learning and suboptimal performance in the identification task.

\begin{figure}[t]
  \centering
  \includegraphics[width=1\linewidth]{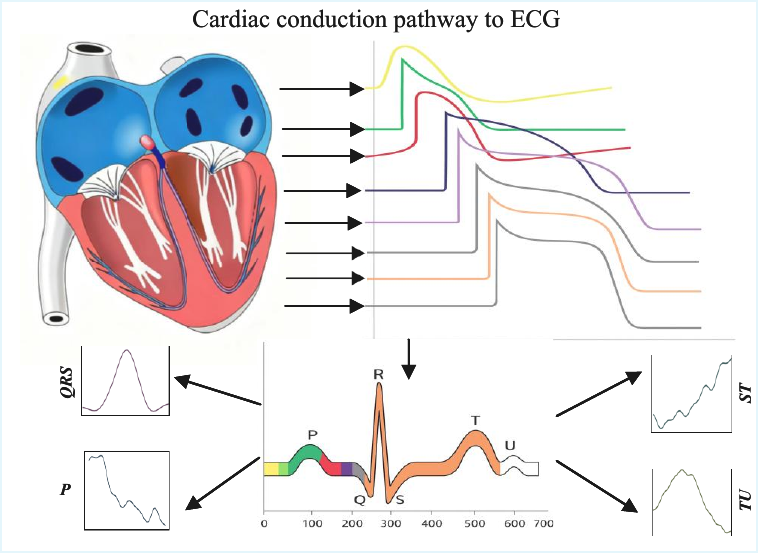}
  \caption{An illustration of the relationship between the different phases of the cardiac cycle and the corresponding ECG signal.}
  \label{fig:concept}
\end{figure}

Specifically, the ECG exhibits distinct waveform phases that originate from different electrophysiological activities of the heart, including the P wave, QRS complex, and T wave. Due to their distinct electrophysiological mechanisms, these phases exhibit markedly different morphological and spectral characteristics, as illustrated in Fig.~\ref{fig:concept}. The P wave reflects atrial depolarization with relatively smooth and low-amplitude morphology, while the QRS complex corresponds to rapid ventricular depolarization and presents the highest energy and steepest transitions due to the large myocardial mass and fast conduction velocity. The T wave represents ventricular repolarization, characterized by slower electrical recovery and smoother waveform patterns. Hence, treating these physiologically distinct phases as homogeneous signal patterns would severely limit the model’s ability to capture stable and discriminative identity features.

Inspired by the intrinsic physiological structure of cardiac cycles, we propose a multi-granularity cardiac-phase representation framework that phase-aware models the heterogeneous phase patterns.
Specifically, at first, we propose a Cardiac Phase Segmentation~(CPS) method to segment each heartbeat into four distinct phases, including P, QRS, ST, and T/U, which is anchored at the R-peak. In this way, our method obtains physiologically meaningful temporal dynamics phases and preserves phase-specific morphological characteristics, providing a reliable basis for subsequent phase-specific modeling.

As previously discussed, employing a single shared encoder ignores the physiological heterogeneity across phases, hindering stable and discriminative representation extraction. To address this issue, we propose a Hierarchical Phase-Aware Fusion strategy~(HPAF), including the Intra-Phase Representation~(IPR), Phase-Grouped Hierarchical Fusion~(PGHF), and Global Representation Fusion~(GRF) modules. In IPR, to avoid inter-phase feature entanglement, we design phase-specific Morphology–Variation-aware Feature Extractors (MVFEs), where each phase is processed by an independent MVFE to extract discriminative features. MVFE consists of a standard convolution branch for morphology extraction and a learnable Gabor branch for variation modeling. The standard convolution branch can effectively capture the waveform morphology and temporal structures, while the learnable Gabor filters are capable of modeling fine-grained local variations such as subtle oscillations and abrupt transitions.

Then, in PGHF, we divide the phases into two physiologically coherent groups and perform phase-grouped hierarchical fusion separately: one PGHF module for the P and T/U features, and another for the QRS and ST features. Specifically, the P and T/U features are grouped as they are both related to repolarization and exhibit smooth, low-amplitude waveforms. In contrast, the QRS and ST features are grouped due to their association with ventricular depolarization and early repolarization, characterized by higher amplitude, sharper transitions, and more complex morphologies. Then, we design a cross-phase gated fuser that selectively couples phases through gated calibration rather than directly fusing them, enabling stable calibration for different phase groups.

Finally, the two features fused by the PGHF modules are fed into the GRF module, where a cross-phase gated fuser is applied again to adaptively fuse them into the final representation. Concretely, GRF first projects each fused feature into a shared scoring space and computes a scalar attention-like gate that measures the relative importance of the QRS–ST group versus the ST–T/U group. This gate then controls a weighted mixture of the two fused features, followed by a linear projection to obtain the final global feature. In this way, the model can dynamically decide whether fast transient cues (QRS–ST) or slow repolarization cues (ST–T/U) should dominate under different noise conditions and rhythm patterns.

To optimize our network and extract discriminative features, we introduce a contrastive loss. This loss constrains global fused embeddings in the feature space to encourage balancing the phase-specific identity cues and cross-phase consistency. In the implementation stage, since ECG signals are continuously acquired, multiple heartbeats can be collected for each individual. We propose a Heartbeat-Aware Multi-prototype~(HAM) enrollment strategy. Unlike previous works that represent each individual with a single feature vector, HAM represents each subject using a set of prototypes. In this way, our method can obtain a gallery set that improves the matching reliability and the performance through reducing the impact of heartbeat-specific noise and variability. Our main contributions can be summarized as follows:

\begin{itemize}
  \item We propose HPAF, a physiologically grounded phase-aware ECG identification framework that avoids cross-phase feature entanglement through a three-stage design.
  
  \item We design a dual-branch MVFE to capture the morphology and variation features, and introduce PR-GAT to model their interactions for reliability-aware refinement explicitly.

  \item We propose a novel enrollment strategy HAM, which leverages multiple heartbeats within ECG signals to construct a robust multi-prototype gallery template for each individual.
  \item Extensive experiments on the public datasets validate the effectiveness through comprehensive experiments under closed-set and open-set protocols.
\end{itemize}


\section{Related Works}
\subsection{Representation Learning for ECG Identification}
\subsubsection{Fiducial-Based Feature Learning}
Fiducial-based ECG identification methods rely on the precise detection of salient landmarks (P wave, QRS complex, T wave) and construct time-domain features. For example, Biel \textit{et al.}~\cite{2} proposed to extract amplitudes, durations and inter-wave intervals from 12-lead resting ECG and use a SIMCA classifier for subject verification. Israel \textit{et al.}~\cite{3} derive normalized temporal features from carefully annotated P/R/T points on high-rate ECG and combine LDA with majority voting to achieve robust recognition under varying electrode locations and anxiety states. Accurate delineation is therefore a key prerequisite for fiducial pipelines; for instance, Martínez \emph{\textit{et al.}}~\cite{WT_Delineator} developed a wavelet-transform-based ECG delineator to localize QRS complexes and delineate P/QRS/T peaks and boundaries on standard databases. These landmark-based schemes are physically interpretable but highly sensitive to boundary detection errors, motion and EMG noise, as summarized by Fratini \emph{\textit{et al.}}~\cite{12}, whereas Wang \emph{\textit{et al.}}~\cite{13}. wavelet–based time–frequency representation with stacked sparse autoencoders dispenses with explicit fiducial detection, underscoring the limitations of traditional fiducial approaches in noisy and cross-condition scenarios.
\subsubsection{Non-Fiducial Representation Learning}
Non-fiducial feature extraction bypasses explicit detection of P/QRS/T landmarks and instead derives descriptors from the whole ECG or its transformed/reconstructed domains. Representative methods include Jung \emph{\textit{et al.}}~\cite{14}, who segment preprocessed single-lead ECG into fixed windows and compute autocorrelation- and DCT-based features with a window-removal strategy and classical classifiers; Fang \emph{\textit{et al.}} ~\cite{15}, who reconstruct ECG phase space and quantify similarity/dissimilarity between portraits; Srivastva \emph{\textit{et al.}}~\cite{16}, who combine autocorrelation with DCT/DFT/WHT transforms and PCA/LDA; and Goshvarpour \emph{\textit{et al.}}~\cite{Goshvarpour2019}, who use Matching Pursuit–based statistical and nonlinear features with PNN/kNN and feature selection. Gutta \emph{\textit{et al.}}~\cite{GuttaCheng}, who reformulate multi-class ECG identification as one-vs-all multi-task learning and use a feature-scaled probabilistic kernel classifier with sparse--low-rank weight decomposition for joint feature selection and classification.These approaches avoid errors caused by inaccurate fiducial detection, but typically produce high-dimensional representations that require feature selection or dimensionality reduction.
\subsubsection{Deep learning–based feature extraction}
In recent years, deep learning has become the mainstream solution for ECG-based identity recognition, enabling end-to-end learning of discriminative features. Most approaches rely on time-domain CNNs operating on R-peak–anchored beats or rhythm segments, while frequency information is added via fixed time–frequency transforms (STFT, CWT) or handcrafted Gabor descriptors fused a posteriori, which can introduce time–frequency crosstalk and make performance sensitive to window length and sampling rate. Representative examples include single-beat time–frequency modeling around the R peak~\cite{SINGLEBEAT_TF}, 2D CNNs on STFT spectrograms for rhythm classification~\cite{STFT_ECG_CNN}, multi-scale residual CNNs and Siamese architectures for identity verification~\cite{PMS_RESNET, EDITH, RDS_CNN}, and LSTM-based models that capture long-range temporal dependencies across multiple beats.
\subsection{Attention mechanisms for ECG biometrics}
Recent ECG biometric systems increasingly exploit attention to emphasize informative temporal patterns and heartbeats. Chee \emph{\textit{et al.}}~\cite{TransAtt_ECG} employ a Transformer-style self-attention encoder on ECG sequence pairs to learn flexible identity embeddings that support both identification and verification across multiple databases. Jyotishi \emph{\textit{et al.}}~\cite{HierLSTM_Att} design a hierarchical LSTM with an internal attention module that assigns larger weights to ECG complexes carrying richer biometric information, outperforming traditional fiducial and non-fiducial baselines on on-body and off-body datasets. At the convolutional level, Hammad \emph{\textit{et al.}}~\cite{ResNet_Att} integrate an attention block into a residual CNN (ResNet-Attention) so that discriminative ECG segments are highlighted in feature maps, while Pan \emph{\textit{et al.}}~\cite{ADAFN} introduce a multi-branch domain-adaptive fusion network where a weighted adaptive attention mechanism reinforces session-robust heartbeat features on ECGID, PTB, CYBHi, and Heartprint.

Beyond purely temporal attention, several works apply attention to channel, feature, or graph spaces. Sun \emph{\textit{et al.}}~\cite{RandAtt_DualPath} propose RandSaD, a dual-path residual model with split attention that selectively aggregates multi-scale channel features from wavelet-denoised ECG windows and reaches $\approx 99.6\%$ identification accuracy. Wang \emph{\textit{et al.}}~\cite{DualLevel_DimAtt} fuse 1D and 2D ECG representations within a collaborative embedding framework and use dimensional attention to up-weight discriminative latent features. Ma \emph{\textit{et al.}}~\cite{OOD_GNN_ECG} present ORGNNFL, which couples a multi-feature attention module with graph neural network fusion to adaptively aggregate topology-aware ECG features under distribution shifts. Al Alfi \emph{\textit{et al.}}~\cite{GCN_12Lead}construct a 12-lead ECG graph whose edges are weighted by mutual-information-based ``attention-like'' scores.


\section{Proposed Method}
\label{sec:proposed}

\begin{figure*}[!t]
  \centering
  \includegraphics[width=\textwidth]{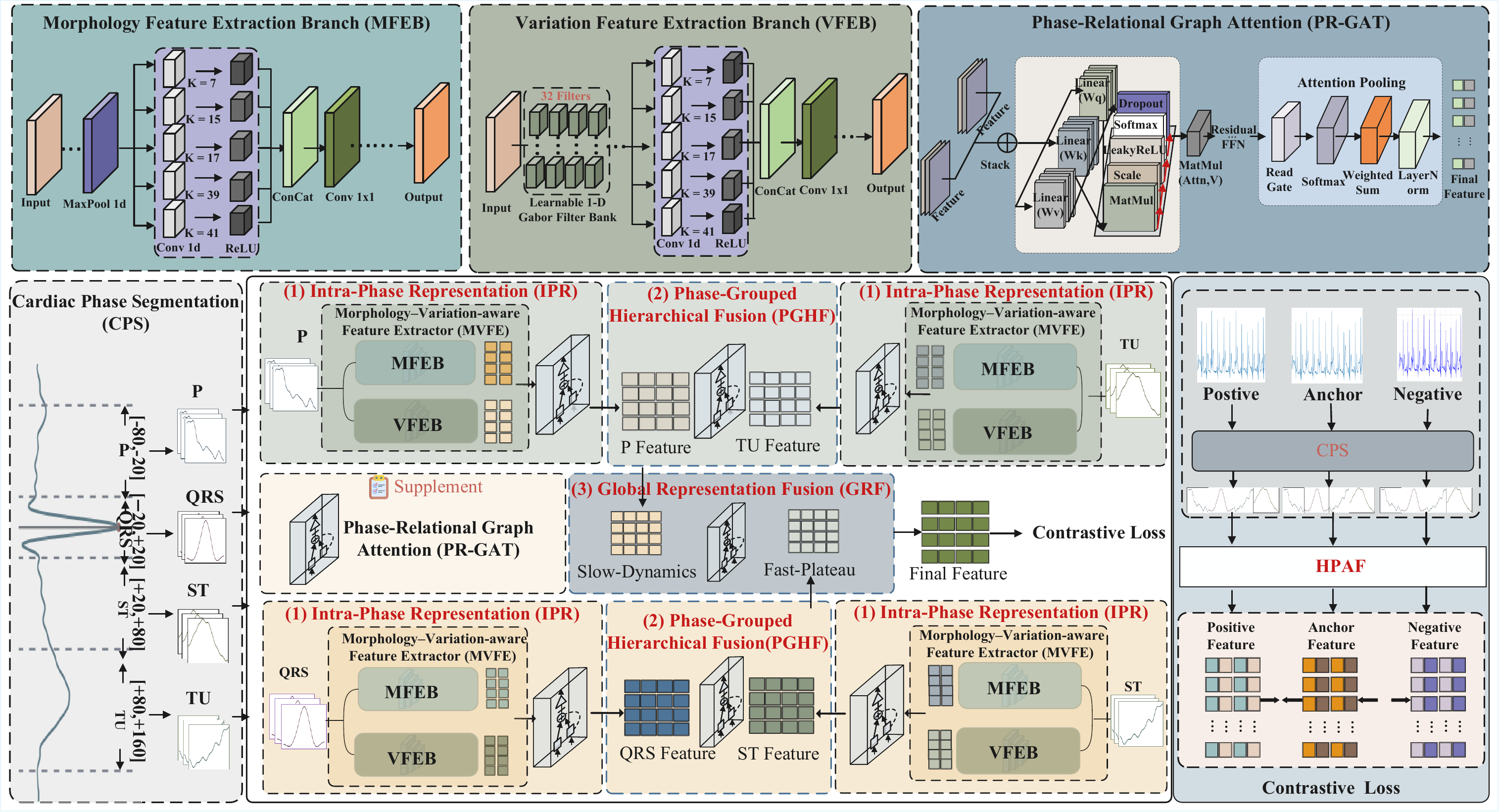}
  \caption{
We present the proposed HPAF framework for phase-aware ECG identification. CPS detects R-peaks and partitions each heartbeat into four phase-aligned segments, namely P, QRS, ST, and T/U. Each phase is encoded by MVFE with MFEB and VFEB, and fused by PR-GAT to form phase representations, which PGHF and GRF then aggregate into a single beat-level identity embedding. The encoder is trained with the contrastive loss to increase inter-subject separability.
}
  \label{fig:overview}
\end{figure*}
\subsection{Overview}

In this paper, we propose a novel multi-granularity cardiac-phase representation framework for ECG identification, and the overview of our proposed method can be found in Figure~\ref{fig:overview}. Employing a single shared encoder to extract features across different cardiac phases would neglect the physiological heterogeneity, which may hinder the extraction of stable and discriminative features.
To mitigate this issue, we propose the Cardiac Phase Segmentation~(CPS) method to segment the ECG into four cardiac phases to avoid phase distortion in the subsequent feature extraction. Then, we propose a Hierarchical Phase-Aware Fusion~(HPAF) strategy. HPAF consists of Intra-Phase Representation (IPR), Phase-Grouped Hierarchical Fusion (PGHF), and Global Representation Fusion (GRF) modules.

Specifically, in IPR, each phase is processed by a phase-specific Morphology–Variation-aware Feature Extractor (MVFE) to avoid inter-phase feature entanglement. Each MVFE contains a convolution branch for waveform morphology and a learnable Gabor branch for fine-grained local variations. PGHF then groups the phases into P–T/U and QRS–ST, and performs hierarchical fusion within each group via a cross-phase gated fuser. GRF then aggregates the two group-level representations into a unified beat-level presentation as the final feature. Moreover, a contrastive loss is further introduced to enhance cross-phase consistency while preventing over-reliance on any single phase. These components are detailed in the following.


\subsection{Hierarchical Phase-Aware Fusion (HPAF)}
\label{sec:hpaf}
\subsubsection{Cardiac Phase Segmentation (CPS)}
To alleviate the heterogeneity across different ECG phases, we propose the CPS method for explicit phase segmentation, which effectively prevents feature entanglement among different phases during subsequent feature extraction.
Specifically, CPS treats the R peak as the anchor to explicitly partition each heartbeat into different physiological phases, instead of the conventional average segmenting of the entire ECG signal. Each heartbeat is divided into four stages, including atrial depolarization, ventricular depolarization and early plateau, and ventricular repolarization, corresponding to the P wave, QRS complex, ST segment, and T/U wave.

Specifically, CPS first detects the R peak in the preprocessed ECG signal and then delineates four time windows around each R peak, which approximately correspond to the P, QRS, ST, and T/U phases based on predefined relative offsets and durations. 
The four fixed-length segments are extracted by selecting samples with predefined offset windows, which are measured in samples relative to the R peak: [–80, –20] for P, [–20, +20] for QRS, [+20, +80] for ST, and [+80, +160] for TU.
In this way, CPS produces four-phase segments with clear physiological meaning and consistent temporal scale, which can effectively prevent feature entanglement during subsequent feature extraction.
each segmented phase is independently encoded using a dedicated phase-specific MVFE.
\subsubsection{Intra-Phase Representation (IPR)}
Following CPS, each segmented phase is independently encoded by a phase-specific MVFE. While commonly used convolutional filters effectively capture coarse morphological information, they often fail to preserve subtle variations in ECG signals. Gabor filters are effective at capturing subtle variations in ECG signals~\cite{28}, as they emphasize localized waveform variations within short temporal regions.
Motivated by these complementary properties, MVFE is designed as a dual-branch architecture, including a CNN-based Morphology Feature Extraction Branch~(MFEB) and a learnable Gabor filter-based Variation Feature Extraction Branch~(VFEB).
Moreover, to model the relationships among morphology and variation representations, we introduce a Phase-Relational Graph Attention~(PR-GAT) module. By treating the morphology and variation representations as graph nodes, PR-GAT explicitly captures their relational dependencies to obtain the comprehensive and discriminative feature.

The VFEB is tailored to extract fine-grained variations. In contrast to conventional Gabor filters with manually specified hyperparameters, we introduce a Learnable Gabor Convolution~(LGC) layer, where the filter parameters are directly from data without requiring prior domain knowledge. The $k$-th channel of the 1D Gabor kernel $g$ at the $t$-th relative position within the kernel is defined as:
\begin{equation}
g_k(t) = \exp\left(-\frac{t^2}{2\sigma_k^2}\right) \cdot \cos(2\pi f_k t + \psi_k),
\label{eq:gabor}
\end{equation}
where $\sigma_k$, $f_k$, and $\psi_k$ are learnable, denoting the scale, center frequency, and phase shift of the Gabor kernel, respectively.
To prevent the model from being influenced by slow-varying signal offsets and to emphasize the local variations, we apply a zero-mean constraint as follows:

\begin{equation}
\hat{g}_k(t) = g_k(t) - \frac{1}{|T|}\sum_{\tau=1}^{T} g_k(\tau),
\label{eq:zero_mean}
\end{equation}
where $T$ is the length of the filter, and $\hat{g}_k$ denotes the zero-mean version of $g_k$.


Then, to simultaneously extract short-term and long-term features, we further design a Multi-Scale Feature Block~(MSFB), which consists of parallel convolutional layers with kernel sizes $k=\{7,15,17,39,41\}$. Then, the extracted multi-scale features are concatenated along the channel dimension, followed by feature fusion using a $1\times1$ convolution and further downsampling. Finally, subsequent feature layers map the representation into a compact embedding $\mathbf{z}_{v}$.

Besides, MFEB focuses on extracting morphological features from ECG signals.
MFEB shares a similar overall architecture with VFEB, with the key difference lying in the first feature extraction stage: MFEB employs conventional convolutional layers, whereas VFEB adopts the LGC layers. Then, through MFEB, the extracted morphological features are represented as $\mathbf{z}_{m}$.



Unlike static concatenation that treats $\mathbf{z}_v$ and $\mathbf{z}_m$ equally, PR-GAT explicitly models interactions between morphology and variation representations via graph attention.

Specifically, we first stack $\mathbf{z}_v$ and $\mathbf{z}_m$ into
$\mathbf{H} \in \mathbb{R}^{2 \times d}$, where $d$ is the feature dimension. Then, we project $\mathbf{H}$ into query, key, and value embeddings with learnable matrices $\mathbf{W}_Q, \mathbf{W}_K, \mathbf{W}_V$ as follows:
\begin{equation}
  \mathbf{Q} = \mathbf{H}\mathbf{W}_Q,\quad
  \mathbf{K} = \mathbf{H}\mathbf{W}_K,\quad
  \mathbf{V} = \mathbf{H}\mathbf{W}_V,
\end{equation}
where $\mathbf{Q}$, $\mathbf{K}$, and $\mathbf{V}$ denote query, key, and value representations, respectively.

Then, the final attention feature $\mathbf{A}$ is obtained through attention-based aggregation as follows:
\begin{equation}
  \mathbf{A}
  = \operatorname{softmax}\!\left(
      \operatorname{LeakyReLU}\!\left(
        \frac{\mathbf{Q}\mathbf{K}^\top}{\sqrt{d_k}}
      \right)
    \right),
\end{equation}
where $\operatorname{LeakyReLU}(\cdot)$ denotes the Leaky ReLu activation function. $d_k$ is the scaling factor.

Then, $\mathbf{A}$ is fed into a standard attention block with residual connections and Layer Norm layers, which yields the refined features $\mathbf{H}' = [\mathbf{h}'_v; \mathbf{h}'_m]$. The two channels of $\mathbf{H}'$ represent the refined variation and morphology features, respectively.


To fuse $\mathbf{h}'_v$ and $\mathbf{h}'_m$, we design one attention pooling module. For
$k \in \{v,m\}$, a lightweight MLP is used to predict a scalar score $s_k$, which is
normalized into a weight. Then, the process can be formulated as:
\begin{equation}
  \alpha_k = \frac{\exp(s_k)}{\sum_{j \in \{v,m\}} \exp(s_j)},\quad k \in \{v,m\},
\end{equation}
where $\alpha$ denotes the normalized feature.

Then, the final embedding can be obtained as follows:
\begin{equation}
  \mathbf{h}_{\text{phase}}
  = \operatorname{LN}\!\left(
      \alpha_v \mathbf{h}'_v + \alpha_m \mathbf{h}'_m
    \right),
\end{equation}
where $\operatorname{LN}(\cdot)$ denotes the Layer Normalization operation.
$\mathbf{h}_{\text{phase}} \in \mathbb{R}^{D}$ is the fused phase representation.
In practice, this mechanism dynamically assigns different weights to different features, thereby enhancing robustness.


\subsubsection{Phase-Grouped Hierarchical Fusion (PGHF)}

Upon obtaining the disentangled phase-specific representations from IPR modules, PGHF is designed to fuse these features based on the physiologically-driven fusion strategy. We first partition the embeddings of four phases into two coherent groups based on their morphological characteristics. Specifically, the P and T/U phases are assigned to the \textit{slow-wave group}, since they exhibit relatively smooth and low-amplitude waveform profiles associated with depolarization-repolarization transitions. Conversely, the QRS and ST phases constitute the \textit{fast-wave group}, distinguished by high-amplitude, sharp transient dynamics driven by ventricular activity. These groups are then processed by the PR-GAT module, which facilitates selective cross-phase coupling via gated calibration, ensuring stable feature adaptation across distinct physiological modes.

\subsubsection{Global Representation Fusion (GRF)} 
To bridge the slow-wave and fast-wave groups into a discriminative embedding, GRF serves as the final stage of our hierarchical architecture. 
Similar to the previous step, we formulate the global fusion as a high-level graph interaction task, rather than naive feature concatenation. Our proposed PR-GAT module treats the embeddings of the two groups as graph nodes and facilitates pairwise interactions to assess the relative importance. This enables our method to dynamically model the relationship between slow and fast dynamics, adaptively synthesizing a holistic beat-level embedding that is robust to inter-beat variability.

\subsection{Contrastive Loss}
\label{sec:pcl}

Verification differs fundamentally from classification in that it involves a matching process between a query code and the gallery template, rather than directly assigning a class label.
Hence, a conventional classification loss is insufficient for verification tasks, as it mainly encourages features to cluster around class centers.
Moreover, such a loss tends to cause overfitting, resulting in poor generalization to unseen identities, which is a common challenge in verification scenarios. This loss fails to adequately constrain the feature space, making it ill-suited for ECG biometrics, where subjects are represented by only a few enrolled beats and must be distinguished from many highly similar negative individuals.

To alleviate this issue, we optimize the encoder with a margin-based contrastive loss rather than the classification loss. For $i$-th sample in the mini-batch, we use its paired positive $P$ (same individual) and mine the negative sample $N$ from the candidate pool consisting of all anchors and positives in the mini-batch. $N$ is selected based on the highest cosine similarity (excluding the anchor and its paired positive). Let $s(\cdot,\cdot)$ denote the cosine similarity formulation. Then, our loss can be formulated as follows:
\begin{equation}
\mathcal{L}_{\mathrm{CL}}
= \frac{1}{B}\sum_{i=1}^{B}
\Big[\, m
+ s\!\big({\mathbf{u}}_i,{\mathbf{u}}_i^{\,N}\big)
- s\!\big({\mathbf{u}}_i,{\mathbf{u}}_i^{\,P}\big)
\,\Big]_+,
\end{equation}
where $m$ is the margin , $[\cdot]_+=\max(\cdot,0)$.
and ${\mathbf{u}}_i$ is denoted the beat-level  feature produced by GRF.
\subsection{Implementation}
\label{sec:implementation}
Since ECG signals are continuously acquired, multiple heartbeats can be collected for each individual. Then, we adopt a Heartbeat-Aware Multi-prototype~(HAM) Enrollment strategy to ensure the verification robustness. Unlike the previous works, they represent each individual via the single feature vector, HAM represents each subject with a set of prototypes. 

The embedding set of the $s$-th individual can be formulated as $\mathcal{E}_{s}=\{\mathbf{u}_{s,1}, ...,\mathbf{u}_{s,N_s}\}$, where $N_s$ denotes the number of the heartbeats. To construct a multi-prototype representation, we cluster $\mathcal{E}_{s}$
into $K$ groups and represent each group by a prototype.

Specifically, the $k$-th prototype $\mathbf{p}_{s,k}$ is computed as the mean
of the embeddings assigned to the $k$-th cluster:
\begin{equation}
\mathbf{p}_{s,k} = \frac{\sum_{n=1}^{N_s}\mathbb{I}\!\left(c_{s,n}=k\right)\mathbf{u}_{s,n}}
{\sum_{n=1}^{N_s}\mathbb{I}\!\left(c_{s,n}=k\right)},
\quad k=1,\ldots,K,
\end{equation}
where $c_{s,n}\in\{1,\ldots,K\}$ denotes the cluster assignment of the $n$-th embedding, and $\mathbb{I}(\cdot)$ is the indicator function.

During verification, each query embedding is matched against all prototypes stored in the gallery. The identity associated with the prototype that yields the minimum distance to the query is assigned as the final matching result. This design helps reduce the impact of noise and bias arising from individual heartbeats, thereby improving the robustness of the verification process.

\section{Experiments}
\label{sec:experiments}

\captionsetup[subfigure]{font=footnotesize,aboveskip=1pt,belowskip=1pt}
\newlength{\cmcrowheight}
\setlength{\cmcrowheight}{0.21\textheight}
\newlength{\rocgridheight}
\setlength{\rocgridheight}{0.22\textheight}
\newcommand{\rocsep}{\hspace{0.012\textwidth}}

\begin{table*}[!t]
  \centering
  \caption{\textbf{Closed-set segment-level identification} across three datasets. Our method consistently outperforms state-of-the-art baselines, demonstrating superior representation learning under intra-subject conditions.}
  \label{tab:ecg_cvpr_table}
  \begin{tabular}{
    l
    c c c
    c c c
    c c c
  }
  \toprule
  \multirow{2}{*}{\textbf{Method}} &
  \multicolumn{3}{c}{\textbf{ECGID}} &
  \multicolumn{3}{c}{\textbf{MIT-BIH}} &
  \multicolumn{3}{c}{\textbf{PTB}} \\
  \cmidrule(lr){2-4}\cmidrule(lr){5-7}\cmidrule(lr){8-10}
  & {\textbf{Acc (\%)}~$\uparrow$} & {\textbf{AUC (\%)}~$\uparrow$} & {\textbf{EER (\%)}~$\downarrow$}
  & {\textbf{Acc (\%)}~$\uparrow$} & {\textbf{AUC (\%)}~$\uparrow$} & {\textbf{EER (\%)}~$\downarrow$}
  & {\textbf{Acc (\%)}~$\uparrow$} & {\textbf{AUC (\%)}~$\uparrow$} & {\textbf{EER (\%)}~$\downarrow$} \\
  \midrule
  PMS\_RESNET &
    96.8700\% & 93.1958\% & 12.4092\% &
    96.9625\% & 99.8303\% &  1.2697\% &
    94.3336\% & 93.1958\% & 12.4092\% \\
  ESN &
    82.9200\% & 85.0162\% & 21.2550\% &
    95.4948\% & 98.5032\% &  4.9168\% &
    92.1900\% & 92.1200\% & 14.3600\% \\
  BAED &
    89.2708\% & 90.1200\% & 13.8700\% &
    98.7302\% & 99.9106\% &  0.7120\% &
    98.0616\% & 93.6600\% & 11.9300\% \\
  ADAFN &
    83.6000\% & 90.6300\% & 13.4800\% &
    96.8900\% & 99.7700\% &  0.9200\% &
    92.9600\% & 96.5300\% &  9.7000\% \\
  EDITH &
    91.8700\% & 90.1500\% & 14.5600\% &
    95.6700\% & 94.0900\% & 10.8000\% &
    98.8500\% & 96.1100\% &  8.3800\% \\
  SSL\_CNN &
    88.4347\% & 94.2343\% & 12.1726\% &
    98.7169\% & 12.1726\% &  0.4241\% &
    98.7069\% & 92.9626\% & 13.5638\% \\
  RDS\_CNN &
    88.3745\% & 90.5300\% & 14.9500\% &
    98.6600\% & 99.1208\% &  4.0864\% &
    92.5735\% & 90.8400\% & 15.8600\% \\
  \textbf{Ours} &
    \bestnum{97.4829\%} & \bestnum{95.4247\%} & \bestnum{9.9058\%} &
    \bestnum{99.7553\%} & \bestnum{99.9692\%} & \bestnum{0.1114\%} &
    \bestnum{99.1678\%} & \bestnum{96.8652\%} & \bestnum{8.6857\%} \\
  \bottomrule
  \end{tabular}
  \vspace{-4pt}
\end{table*}

\begin{table*}[!t]
  \centering
  \caption{\textbf{Open-set segment-level identification} across ECGID, MIT-BIH, and PTB. Our model shows robust generalization to unseen subjects across datasets with diverse pathologies and acquisition setups.}
  \label{tab:ecg_openset_table}
  \begin{tabular}{
    l
    c c c
    c c c
    c c c
  }
  \toprule
  \multirow{2}{*}{\textbf{Method}} &
  \multicolumn{3}{c}{\textbf{ECGID}} &
  \multicolumn{3}{c}{\textbf{MIT-BIH}} &
  \multicolumn{3}{c}{\textbf{PTB}} \\
  \cmidrule(lr){2-4}\cmidrule(lr){5-7}\cmidrule(lr){8-10}
  & {\textbf{Acc (\%)}~$\uparrow$} & {\textbf{AUC (\%)}~$\uparrow$} & {\textbf{EER (\%)}~$\downarrow$}
  & {\textbf{Acc (\%)}~$\uparrow$} & {\textbf{AUC (\%)}~$\uparrow$} & {\textbf{EER (\%)}~$\downarrow$}
  & {\textbf{Acc (\%)}~$\uparrow$} & {\textbf{AUC (\%)}~$\uparrow$} & {\textbf{EER (\%)}~$\downarrow$} \\
  \midrule
  PMS\_RESNET &
    82.4300\% & 84.6029\% & 22.7084\% &
    89.2250\% & 85.0718\% & 23.7836\% &
    80.7700\% & 82.3452\% & 24.5576\% \\
  ESN &
    71.7800\% & 71.2400\% & 34.3700\% &
    69.4326\% & 66.0120\% & 39.5100\% &
    70.4300\% & 69.5938\% & 35.2144\% \\
  BAED &
    71.2361\% & 69.8344\% & 34.1535\% &
    89.3586\% & 87.3643\% & 19.0133\% &
    85.4745\% & 94.7807\% &  9.6553\% \\
  ADAFN &
    64.6900\% & 65.3800\% & 34.9100\% &
    57.7700\% & 66.6600\% & 37.9600\% &
    60.5900\% & 63.6300\% & 38.0400\% \\
  EDITH &
    87.8700\% & 92.2100\% & \bestnum{12.4300\%} &
    96.6600\% & 97.4600\% & \bestnum{6.0400\%} &
    98.6500\% & 95.1100\% & 10.2800\% \\
  SSL\_CNN &
    57.1344\% & 66.9551\% & 38.8534\% &
    90.0811\% & 90.0811\% & 17.8188\% &
    87.8999\% & 87.0397\% & 20.0727\% \\
  RDS\_CNN &
    89.3680\% & 83.1503\% & 22.1800\% &
    91.4315\% & 91.0700\% & 15.9090\% &
    94.5760\% & 91.6559\% & 15.5920\% \\
  \textbf{Ours} &
    \bestnum{94.4709\%} & \bestnum{92.5548\%} & 15.1465\% &
    \bestnum{98.2179\%} & \bestnum{97.6833\%} &  8.0360\% &
    \bestnum{98.9339\%} & \bestnum{96.2483\%} & \bestnum{9.6234\%} \\
  \bottomrule
  \end{tabular}
  \vspace{-4pt}
\end{table*}

\begin{figure*}[!t]
  \centering
  
  \begin{subfigure}[t]{0.33\textwidth}
    \centering
    \includegraphics[width=\linewidth,height=\cmcrowheight,keepaspectratio]{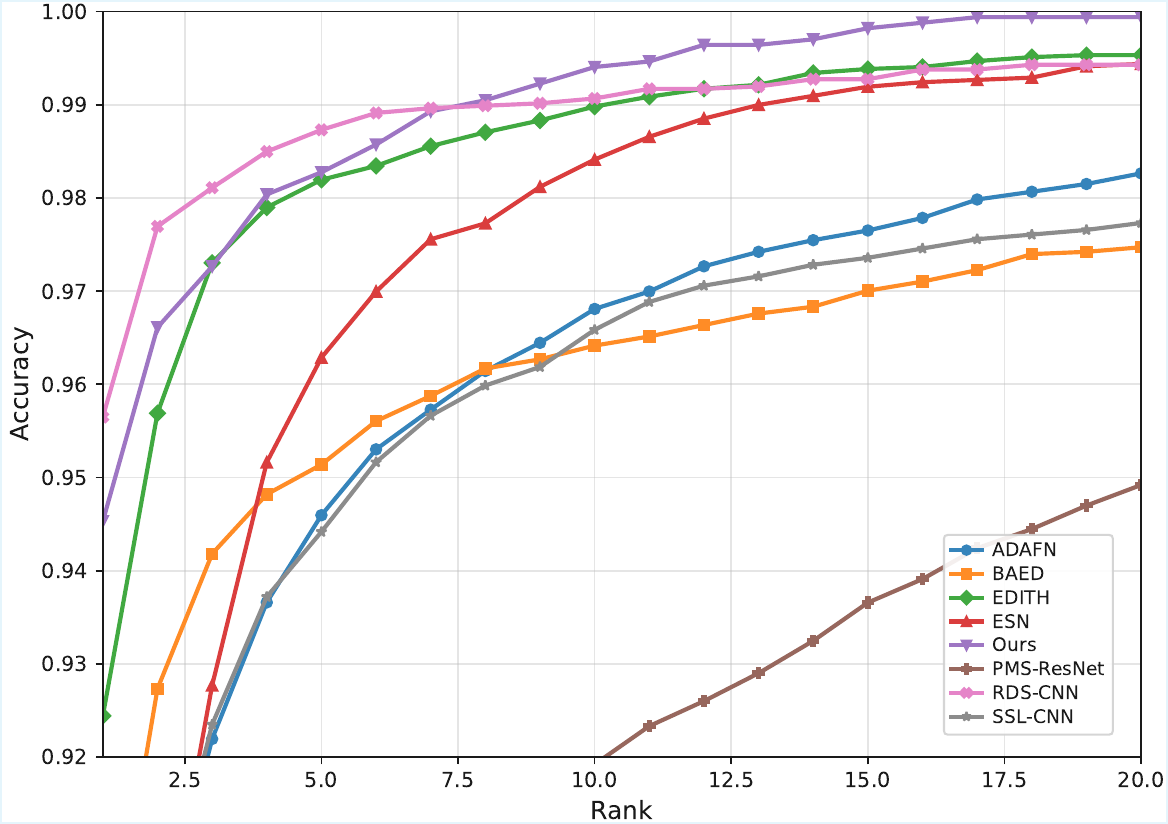}
    \caption{ECGID}
    \label{fig:cmc_closed_ecgid}
  \end{subfigure}\hfill
  \begin{subfigure}[t]{0.33\textwidth}
    \centering
    \includegraphics[width=\linewidth,height=\cmcrowheight,keepaspectratio]{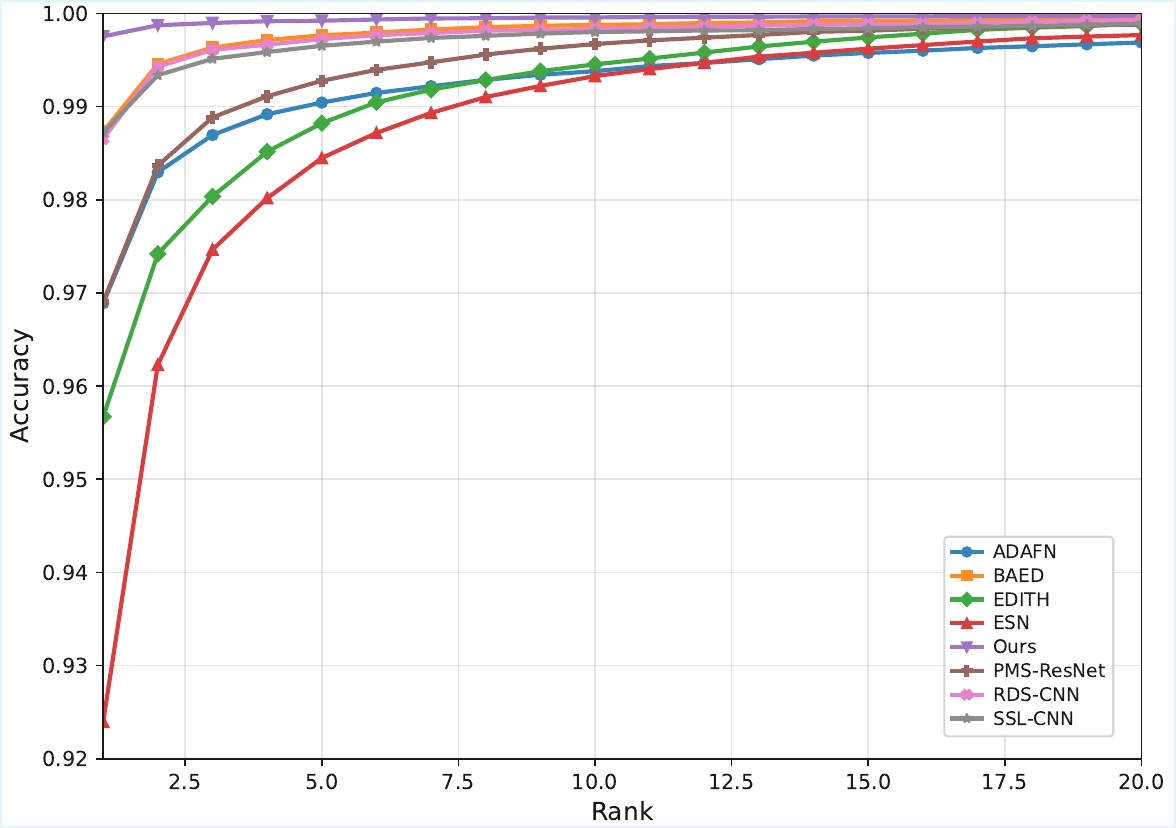}
    \caption{MIT-BIH}
    \label{fig:cmc_closed_mit}
  \end{subfigure}\hfill
  \begin{subfigure}[t]{0.33\textwidth}
    \centering
    \includegraphics[width=\linewidth,height=\cmcrowheight,keepaspectratio]{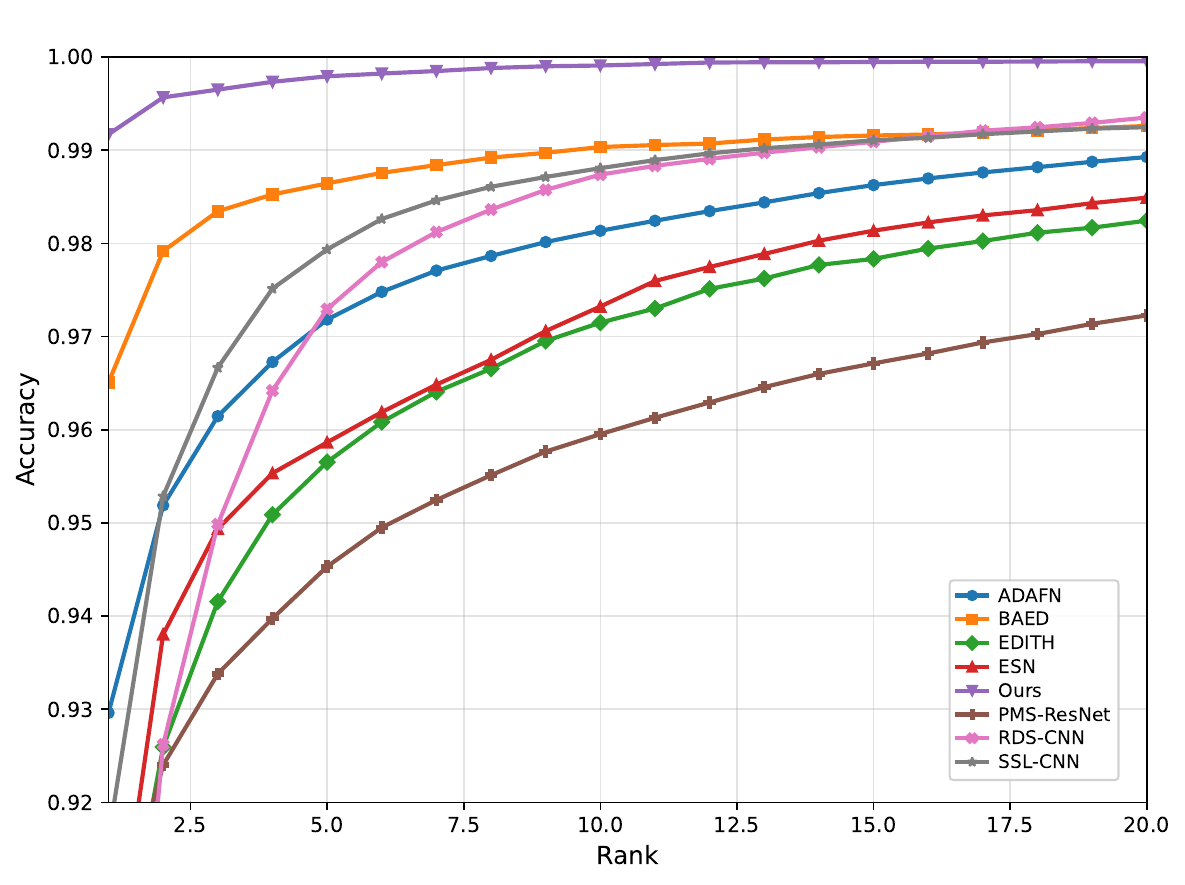}
    \caption{PTB}
    \label{fig:cmc_closed_ptb}
  \end{subfigure}

  \vspace{4pt}

  \begin{subfigure}[t]{0.33\textwidth}
    \centering
    \includegraphics[width=\linewidth,height=\cmcrowheight,keepaspectratio]{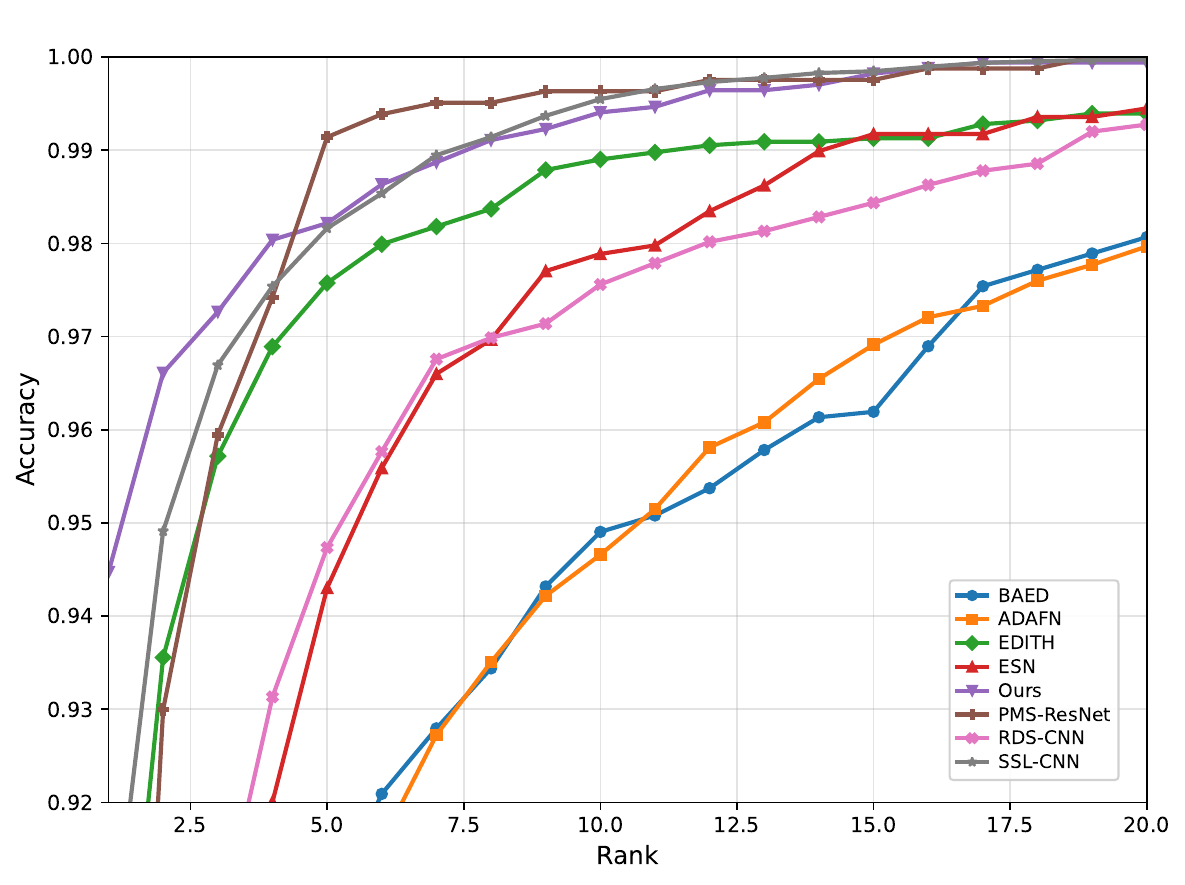}
    \caption{ECGID}
    \label{fig:cmc_open_ecgid}
  \end{subfigure}\hfill
  \begin{subfigure}[t]{0.33\textwidth}
    \centering
    \includegraphics[width=\linewidth,height=\cmcrowheight,keepaspectratio]{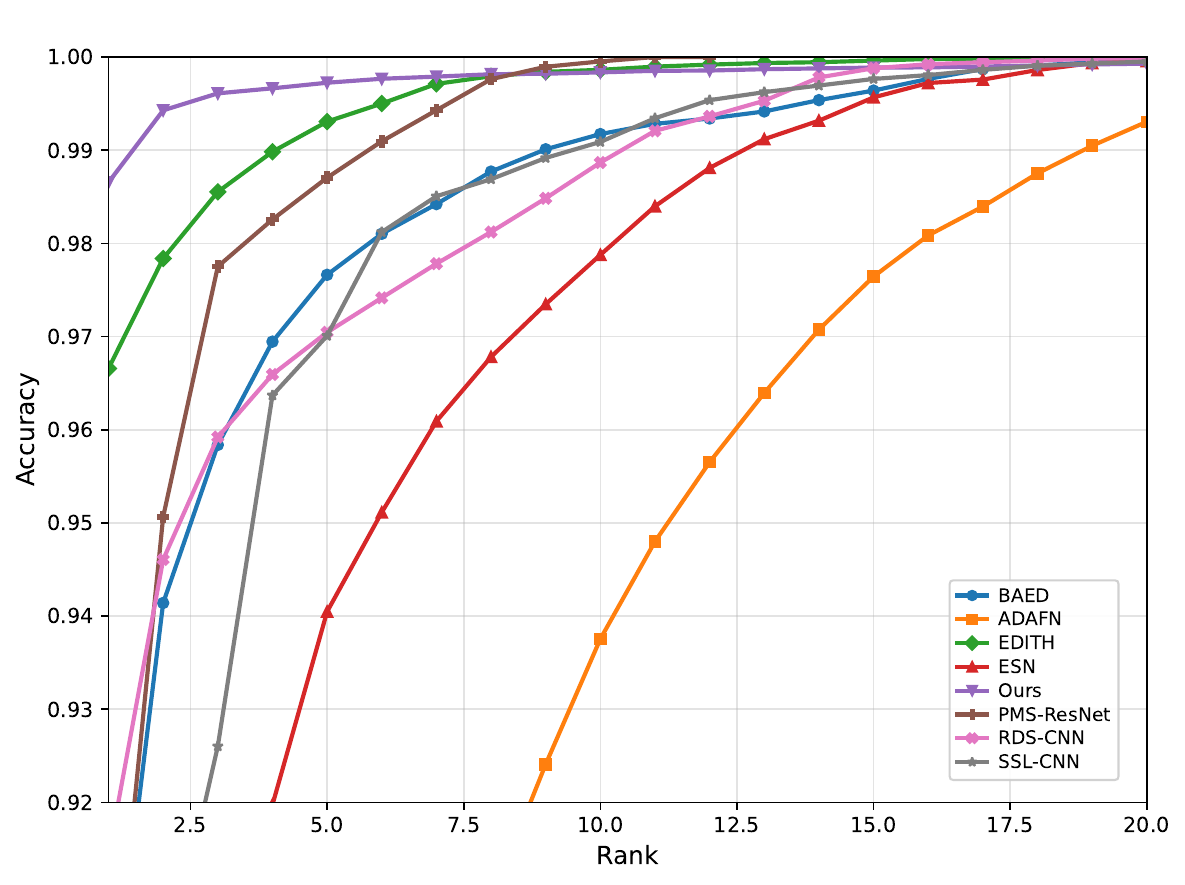}
    \caption{MIT-BIH}
    \label{fig:cmc_open_mit}
  \end{subfigure}\hfill
  \begin{subfigure}[t]{0.33\textwidth}
    \centering
    \includegraphics[width=\linewidth,height=\cmcrowheight,keepaspectratio]{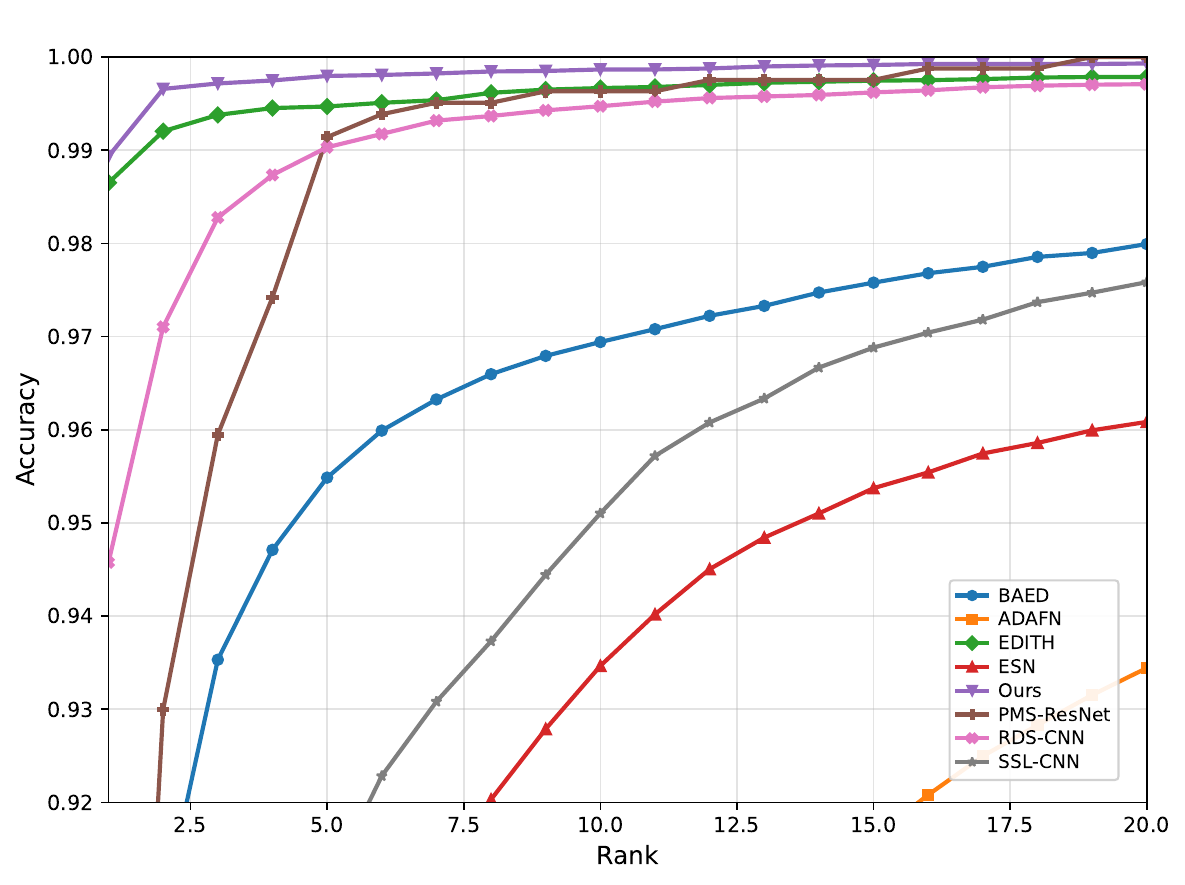}
    \caption{PTB}
    \label{fig:cmc_open_ptb}
  \end{subfigure}

  \caption{CMC comparisons of different methods across three datasets.
Subplots~(\subref{fig:cmc_closed_ecgid})--(\subref{fig:cmc_closed_ptb}) denote the performance of different methods on the
MIT-BIH, ECGID, and PTB datasets under the closed-set setting,
respectively, while~(\subref{fig:cmc_open_ecgid})--(\subref{fig:cmc_open_ptb}) show the corresponding results under the
open-set setting.}
  \label{fig:cmc_3x2_rows}
\end{figure*}
\begin{table*}[t]
\centering
\caption{Ablation on basic segmentation with two branches across datasets.}
\label{tab:ablation_cnn_gabor_all}
\resizebox{\textwidth}{!}{
\begin{tabular}{llccc ccc ccc}
\toprule
\multirow{2}{*}{Protocol} & \multirow{2}{*}{Method}
& \multicolumn{3}{c}{ECGID} & \multicolumn{3}{c}{MIT-BIH} & \multicolumn{3}{c}{PTB} \\
\cmidrule(lr){3-5}\cmidrule(lr){6-8}\cmidrule(lr){9-11}
& & Acc(\%)$\uparrow$ & AUC(\%)$\uparrow$ & EER(\%)$\downarrow$
  & Acc(\%)$\uparrow$ & AUC(\%)$\uparrow$ & EER(\%)$\downarrow$
  & Acc(\%)$\uparrow$ & AUC(\%)$\uparrow$ & EER(\%)$\downarrow$ \\
\midrule
\multirow{3}{*}{Closed-set}
& Only-CNN   & 88.4347 & 94.2343 & 12.1726 & 90.9963 & 99.8905 & 2.0421 & 98.7069 & 92.9626 & 13.5638 \\
& Only-Gabor & 87.9113 & 94.7270 & 11.5600 & 99.6345 & 99.6345 & 1.9013 & 98.0266 & 94.6215 & 11.7848 \\
& Ours       & \textbf{97.4829} & \textbf{95.4247} & \textbf{9.9058} & \textbf{99.7553} & \textbf{99.9692} & \textbf{0.1114} & \textbf{99.1678} & \textbf{96.8952} & \textbf{8.6857} \\
\midrule
\multirow{3}{*}{Open-set}
& Only-CNN   & 57.1344 & 66.9551 & 38.8534 & 90.0811 & 90.0811 & 17.8188 & 87.8999 & 87.0397 & 20.0727 \\
& Only-Gabor & 67.5981 & 79.7807 & 27.7885 & 85.3700 & 88.8950 & 19.3803 & 88.7787 & 89.0093 & 19.3619 \\
& Ours       & \textbf{94.4709} & \textbf{92.5548} & \textbf{15.1465} & \textbf{98.2179} & \textbf{97.6833} & \textbf{8.0360} & \textbf{98.9339} & \textbf{96.2483} & \textbf{9.6234} \\
\bottomrule
\end{tabular}}
\end{table*}

\begin{table}[t]
\centering
\caption{Ablation on staged module stacking (IPR is always enabled).}
\label{tab:ablation_checkmark_allinone}
\setlength{\tabcolsep}{7pt}
\renewcommand{\arraystretch}{1.05}
\resizebox{\columnwidth}{!}{
\begin{tabular}{lccccc}
\toprule
Dataset & CPS & PGHF & GRF & AUC(\%)$\uparrow$ & EER(\%)$\downarrow$ \\
\midrule
\multirow{2}{*}{ECGID}
  & \cmark & \xmark & \xmark & 92.4538 & 15.4915 \\
& \cmark & \cmark & \cmark & \textbf{92.5548} & \textbf{15.1465} \\
\cmidrule(lr){1-6}
\multirow{2}{*}{MIT-BIH}
  & \cmark & \xmark & \xmark & 97.0113 & 9.7294 \\
& \cmark & \cmark & \cmark & \textbf{97.6833} & \textbf{8.0360} \\
\cmidrule(lr){1-6}
\multirow{2}{*}{PTB}
  & \cmark & \xmark & \xmark & 95.7800 & 10.3812 \\
& \cmark & \cmark & \cmark & \textbf{96.2483} & \textbf{9.6234} \\
\bottomrule
\end{tabular}
}
\end{table}

\begin{figure*}[!t]
  \centering
  \begin{subfigure}[t]{0.325\textwidth}
    \centering
    \includegraphics[width=\linewidth,height=\rocgridheight,keepaspectratio]{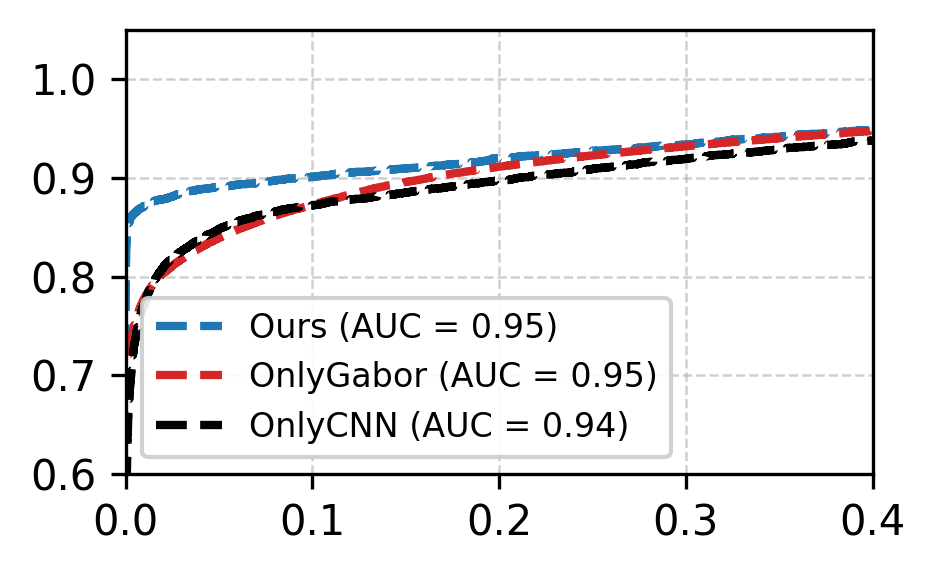}
    \caption{ECGID}
    \label{fig:roc_closed_ecgid}
  \end{subfigure}\rocsep
  \begin{subfigure}[t]{0.325\textwidth}
    \centering
    \includegraphics[width=\linewidth,height=\rocgridheight,keepaspectratio]{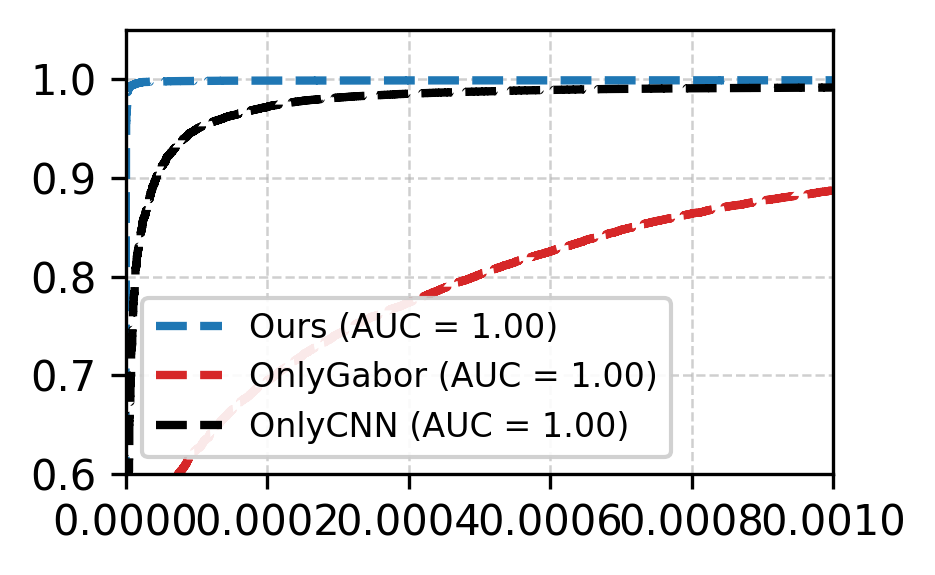}
    \caption{MIT-BIH}
    \label{fig:roc_closed_mit}
  \end{subfigure}\rocsep
  \begin{subfigure}[t]{0.325\textwidth}
    \centering
    \includegraphics[width=\linewidth,height=\rocgridheight,keepaspectratio]{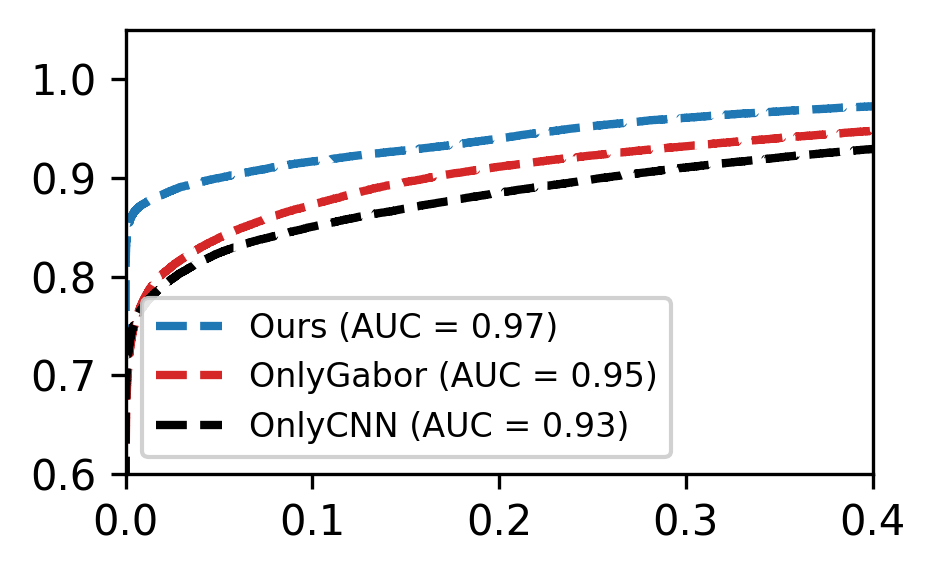}
    \caption{PTB}
    \label{fig:roc_closed_ptb}
  \end{subfigure}

  \vspace{3pt}

  \begin{subfigure}[t]{0.325\textwidth}
    \centering
    \includegraphics[width=\linewidth,height=\rocgridheight,keepaspectratio]{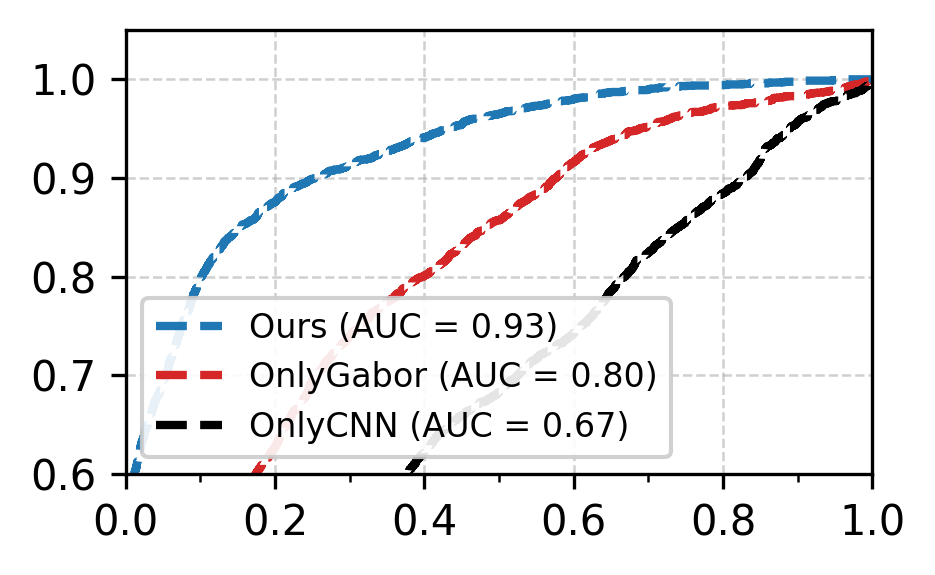}
    \caption{ECGID}
    \label{fig:roc_open_ecgid}
  \end{subfigure}\rocsep
  \begin{subfigure}[t]{0.325\textwidth}
    \centering
    \includegraphics[width=\linewidth,height=\rocgridheight,keepaspectratio]{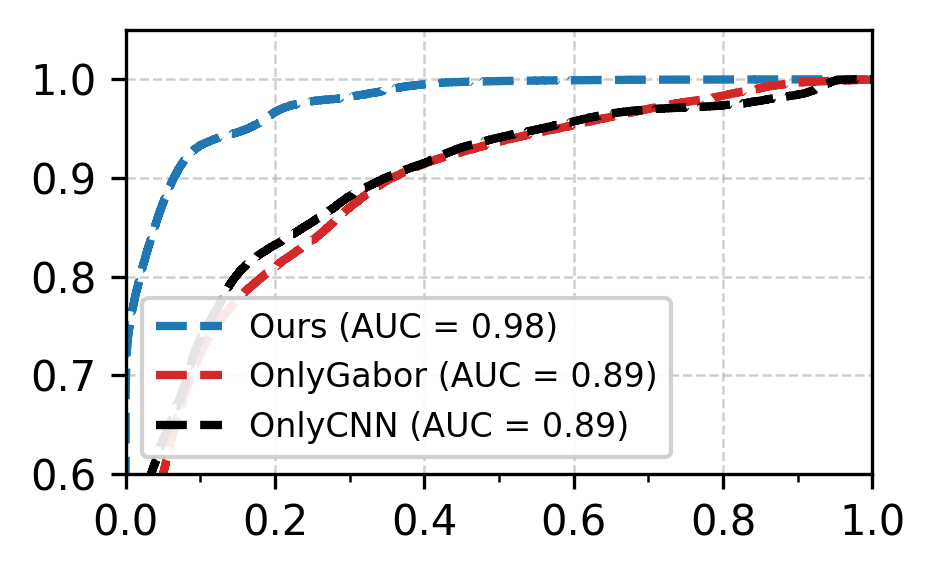}
    \caption{MIT-BIH}
    \label{fig:roc_open_mit}
  \end{subfigure}\rocsep
  \begin{subfigure}[t]{0.325\textwidth}
    \centering
    \includegraphics[width=\linewidth,height=\rocgridheight,keepaspectratio]{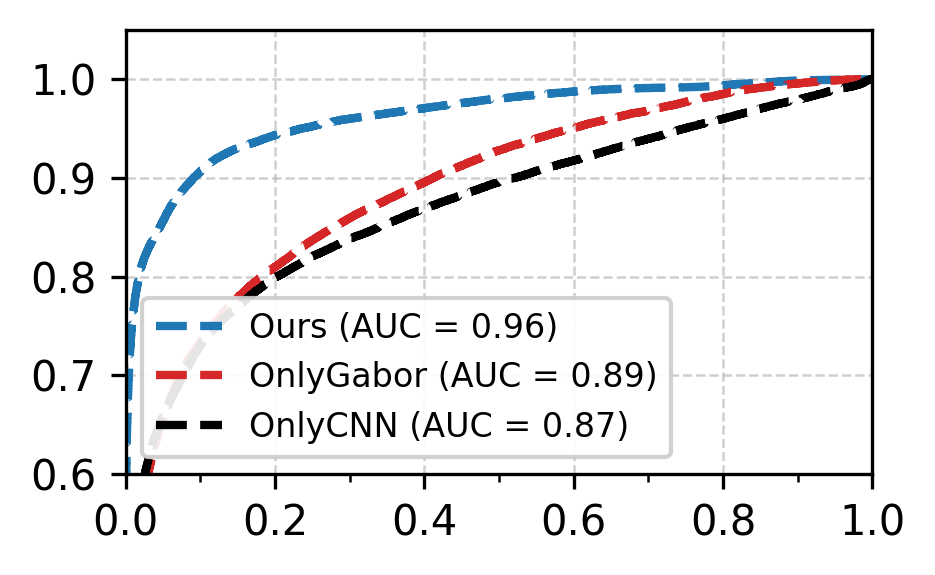}
    \caption{PTB}
    \label{fig:roc_open_ptb}
  \end{subfigure}

  \caption{ROC comparisons of different variants.
  Subplots~(\subref{fig:roc_closed_ecgid})–~(\subref{fig:roc_closed_ptb}) show the ROC curves on ECGID, MIT-BIH, and PTB under
  the closed-set setting, respectively, whereas ~(\subref{fig:roc_open_ecgid})–~(\subref{fig:roc_open_ptb}) depict the
  corresponding results under the open-set setting (same dataset order).}
  \label{fig:roc_rows_3x2}
\end{figure*}

\begin{figure}[t]
  \centering
  \includegraphics[width=1\linewidth]{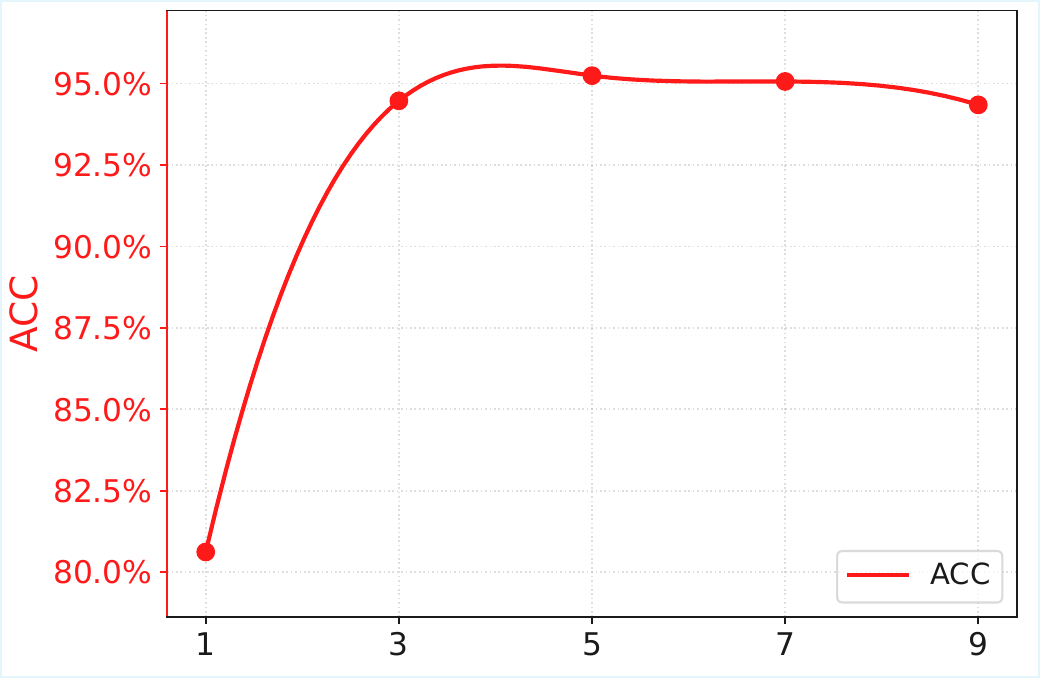}
  \caption{Ablation study about the number of enrollment prototypes in HAM.}
  \label{fig:Effective_of_prototypes}
\end{figure}

\subsection{Datasets and Preprocessing}
\subsubsection{Datasets}
We evaluate our proposed method under different experimental settings in three public ECG datasets.

\noindent\textbf{\textit{ECGID Dataset:}}
Ninety subjects with 310 sessions are provided in the ECGID dataset. Each session contains a 20s single-lead I record with two columns (raw/filtered as column 0/1) sampled at 500Hz, 12-bit. Sessions span 1 day–6 months under unconstrained conditions with coarse R/T annotations~\cite{ECGID_DB}. 

\noindent\textbf{\textit{MIT-BIH Dataset:}}
In the MIT-BIH dataset, forty-eight half-hour Holter excerpts from 47 subjects, typically MLII and V1/V2 leads, at 360\, Hz (11-bit, 10\, mV range) with manually verified beat labels ($\sim$110k beats)~\cite{MITBIH_DB}.

\noindent\textbf{\textit{PTB ECG Dataset:}}
In the PTB dataset, five hundred forty-nine records from 290 subjects (1–5 per subject) with 15 synchronous leads (12 standard + Frank XYZ) at 1000\, Hz/16-bit plus clinical metadata~\cite{PTB_DB}.

\subsubsection{Pre-Processing Details}
To account for variations in lead configuration, sampling rate, and signal quality across ECG datasets, we adopt a unified preprocessing pipeline to standardize model inputs.


The raw signals are denoised using a band-pass filter between 0.5–40Hz to remove baseline drift and high-frequency noise, and are then uniformly resampled to 200Hz. The CPS is applied to detect R-peaks and segment each beat into four phases, including P, QRS, ST, and TU phases. At the unified sampling rate of 200Hz, each sample represents 0.005s. Accordingly, the segment lengths assigned to the P, QRS, ST, and T/U phases are 60, 40, 60 and 80 samples, corresponding to temporal durations of 0.30s, 0.20s, 0.30s, and 0.40s, respectively.


\subsection{Experimental Settings and Comparison Methods}
\subsubsection{Training and Evaluation Settings}

We evaluate our method under both closed-set and open-set settings. In both cases, the data are split at the subject level instead of randomly splitting all samples.


All methods were implemented in PyTorch and trained for 40 epochs using SGD with momentum 0.9, a batch size of 32, and a cosine annealing learning rate schedule initialized at $1{\times}10^{-4}$. Experiments were conducted with an AMD Ryzen 7 5800X CPU and a single NVIDIA RTX 3060Ti GPU.


Similar to the open-set protocol, each query ECG is matched against all enrollment ECGs, and the predicted identity is determined based on the minimum distance criterion.


\subsubsection{Evaluation Metrics and Comparison Methods}

Specifically, we evaluate our method under two tasks, including the verification and identification tasks.  For the verification task, we use Equal Error Rate (EER), defined as the value where the false acceptance and false rejection rates are equal. Besides, Receiver Operating Characteristic (ROC) is also used to evaluate the verification performance. Besides, for the identification task, Top-1 Accuracy (Acc), Area Under the ROC Curve (AUC), and Cumulative Match Characteristic (CMC) curves are used to evaluate the performance. Among them. CMC is used to measure the probability that the correct identity ranks within the top-$k$ predictions.

To ensure fair and comprehensive comparisons, we compare seven baselines, including BAED~\cite{BAED}, RDS-CNN~\cite{RDS_CNN}, EDITH~\cite{EDITH}, PMS-ResNet~\cite{PMS_RESNET}, ADAFFN~\cite{ADAFN}, SSL-CNN~\cite{SSL_CNN}, and ESN~\cite{ESN}. All baselines share the same preprocessing pipeline, input configuration, and training schedule as our method.

\subsection{Closed-Set Experiment}

At first, we evaluate our proposed method under the closed-set setting, where the identities in the training and test sets are identical. For each subject, half of the samples are used for training and the other half for testing.

The quantitative results are listed in Table~\ref{tab:ecg_cvpr_table}, with CMC curves plotted in Fig.~\ref{fig:cmc_3x2_rows}~(\subref{fig:cmc_closed_ecgid})--(\subref{fig:cmc_closed_ptb}). Across all three public datasets, our method yields the highest identification accuracy and lowest EERs. On ECGID, our method achieves 97.48\% accuracy, surpassing both PMS-ResNet (96.87\%) and BAED (89.27\%). The improvement is most evident on the MIT-BIH dataset, where our model records a Top-1 accuracy of 99.75\% and an EER of 0.11\%. Notably, this EER is significantly lower than that of the closest competitors, BAED (0.71\%) and PMS-ResNet (1.27\%). 


\subsection{Open-Set Experiment}

Besides, we evaluate our proposed method under a more challenging setting, the open-set setting, which assumes that the test subjects are unseen during training. All subjects were randomly split into disjoint training and test sets, with model parameters trained exclusively on the training subjects.

For each subject in the test set, half of the samples are used for enrollment, and the remaining half are used as query samples. The feature of each query ECG is matched against all enrollment ECGs' features, and the identity corresponding to the minimum distance is assigned as the final prediction.

Table~\ref{tab:ecg_openset_table} and Fig.~\ref{fig:cmc_3x2_rows}~(\subref{fig:cmc_open_ecgid})--(\subref{fig:cmc_open_ptb}) summarize the results, the general performance decline is observed across baseline methods due to the distribution shift in unseen subjects. SSL-CNN and ADAFN are particularly affected, yielding only 57.13\% and 64.69\% accuracy on ECGID, respectively.
It can be seen that our method is proven robust to this issue. It achieves the best results on all public datasets: 94.47\% accuracy on ECGID (surpassing the second best method by 5.11\%), 98.21\% on MIT-BIH, and 98.93\% on PTB. Besides, we can observe that our method can achieve the lowest EER in almost all settings. The CMC curves reflect this advantage, showing higher identification rates at Rank-1 compared to competing methods.

These results validate the effectiveness and robustness of our strategy, the previous methods achieve satisfactory performance in the closed-set setting but they generally cannot keep it to the unseen individuals. In contrast, our method preserves robust performance by effectively modeling phase-specific information.

\subsection{Ablation Experiments}
As shown in Table~\ref{tab:ablation_cnn_gabor_all} and Fig~\ref{fig:roc_rows_3x2}, we evaluated the effectiveness of each branch in MVFE.
All variants adopt the same NPD segmentation protocol under the same training protocol. It can be noticed that single-branch-based variations exhibit clear limitations in the open-set protocol, while fusing the morphology and variation features improves discrimination across datasets most notably on ECGID. Specifically, AUC increases from 66.96\% (best single-branch) to 92.55\%. These results suggest that the two branches capture complementary cues that are difficult to obtain from a single view alone.

Moreover, to verify the hyperparameter setting about the number of the prototypes in HAM, we validate the performance in the ECGID dataset under the open-set setting, and the results can be found in Fig.~\ref{fig:Effective_of_prototypes}. We observe that the accuracy consistently improves as the number of prototypes increases from 1 to 3, indicating that the proposed multi-prototype design effectively reduces noise introduced by individual heartbeats and leads to more robust performance. However, when the number of prototypes is further increased, the performance tends to saturate and shows no significant variation. Therefore, we empirically recommend using 3 as the number of enrollment prototypes to balance performance and the storage cost.

Finally, the results of the ablation study about the hierarchical fusion strategy can be found in Table IV. It can be noticed that adding PGHF and GRF to the CPS and IPR yields consistent gains, indicating that each module contributes to refining the global embedding and tightening the feature space structure. Importantly, the combined modules achieve the strongest overall performance among the compared variants under the same protocol, supporting the effectiveness of integrating phase-guided hierarchical fusion and global refinement for robust open-set identification.

\section{Conclusion}

In this work, we propose a multi-granularity cardiac-phase representation framework for ECG identification by segmenting each ECG signal into four physiologically meaningful phases. We then propose a hierarchical phase-aware feature extraction framework, HPAF, which consists of the IPR, PGHF, and GRF modules. Specifically, phase-specific MVFEs are assigned to each individual phase to extract phase-specific representations in IPR, after which similar phases are grouped in PGHF and global features are progressively extracted in GRF. Our method achieves competitive performance on public datasets.
Our results suggest that representing an ECG signal as a composition of phase-specific cues, rather than a homogeneous whole-beat pattern, can improve the robustness of ECG-based identity. Though our method achieves impressive performance, a current limitation is that our method relies on the accurate R-peak localization, which may degrade under severe noise or ectopic beats. In our future work, we will explore adaptive phase localization and R-peak-free segmentation methods to achieve more robust performance.

\begin{IEEEbiography}[{\includegraphics[width=1in,height=1.25in, clip,keepaspectratio]{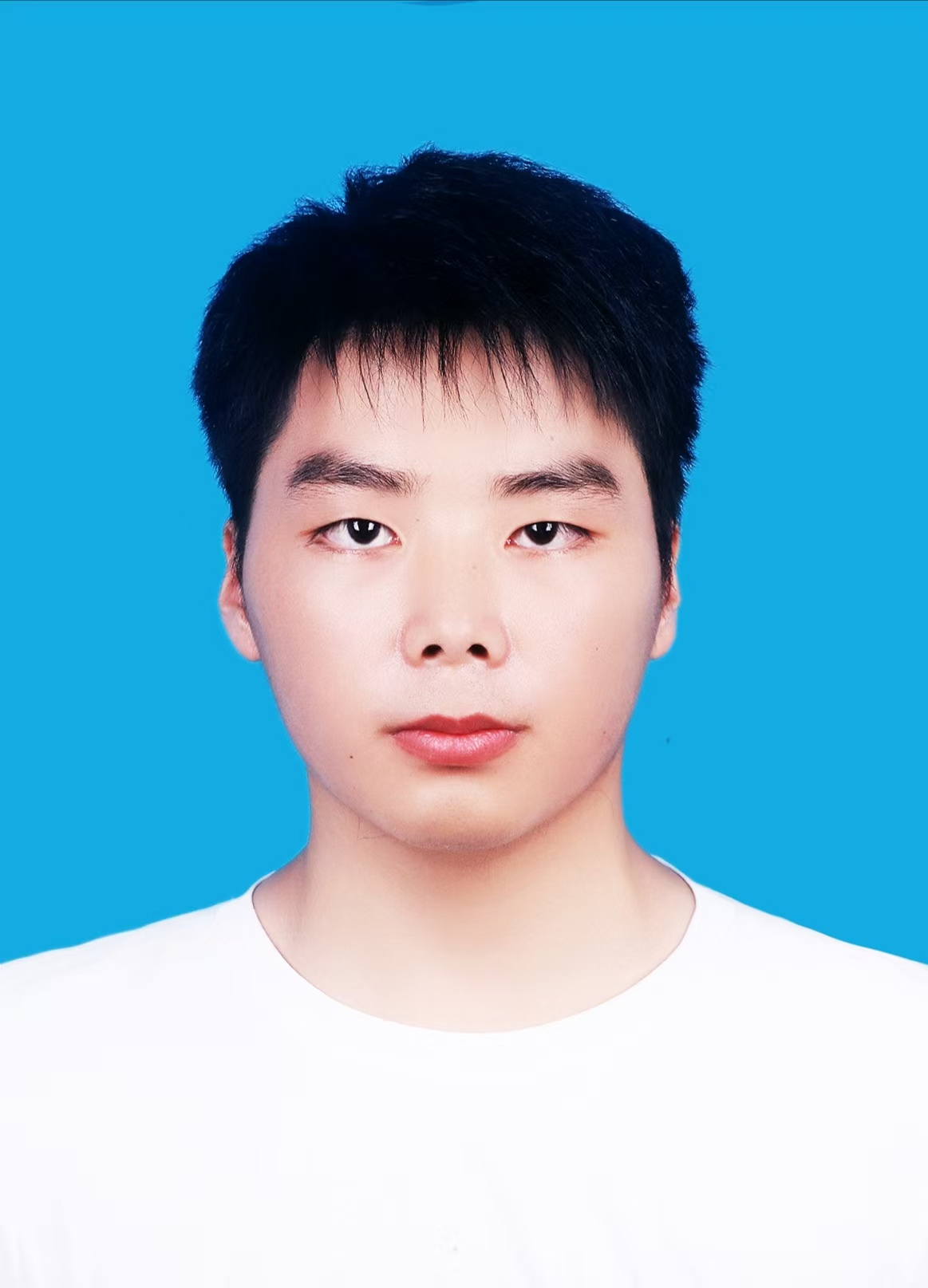}}]{First A.JinTao Huang} received his Bachelor of Engineering degree from the School of Software, Nanchang Hangkong University in 2025. He is currently studying for a master’s degree at Nanchang Hangkong University, China.

His research interests include computer vision, biometric recognition, and applications of generative
artificial intelligence.
\end{IEEEbiography}

\begin{IEEEbiography}[{\includegraphics[width=1in,height=1.25in, clip,keepaspectratio]{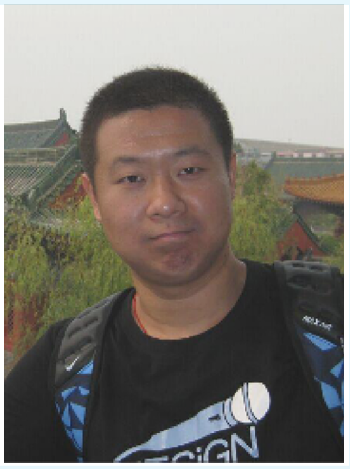}}]{Second B.Lu Leng} received his Ph.D degree
from Southwest Jiaotong University, Chengdu, P. R.China, in 2012. He performed his postdoctoral research at Yonsei University, Seoul, South Korea, and Nanjing University of Aeronautics and Astronautics, Nanjing, P. R. China. He was a visiting scholar at West Virginia University, USA, and Yonsei University, South Korea. Currently, he is a full professor and the dean of Institute of Computer Vision at Nanchang Hangkong University.

Prof. Leng has published more than 100 international journal and conference papers. He has been granted several scholarships and funding projects, including six projects supported by National Natural Science Foundation of China (NSFC). He serves as a reviewer of more than 100 international journals and 50 conferences. His research interests include computer vision, biometric template protection, biometric recognition, medical
image processing, data hiding, etc. 
\end{IEEEbiography}

\begin{IEEEbiography}[{\includegraphics[width=1in,height=1.25in, clip,keepaspectratio]{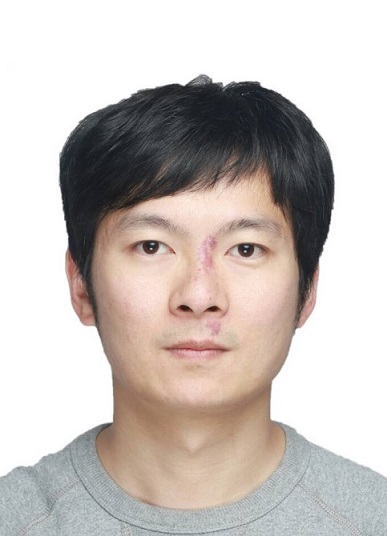}}]{Third C.Yi Zhang} received the B.S., M.S., and Ph.D. degrees in computer science and technology from the College of Computer Science, Sichuan University, Chengdu, China, in 2005, 2008, and 2012, respectively. 

From 2014 to 2015, he was with the Department of Biomedical Engineering, Rensselaer Polytechnic Institute, Troy, NY, USA, as a Postdoctoral Researcher. He is currently a Full Professor with the School of Cyber Science and Engineering, Sichuan University, and is also the Director with the Deep Imaging Group (DIG). He authored more than 140 papers in the field of image processing. These papers were authored or co-authored in several leading journals and conferences, including IEEE Transactions on Medical Imaging, IEEE Transactions on Information Forensics and Security, MedIA, IJCV, and CVPR; and reported by the Institute of Physics (IOP) and during Lindau Nobel Laureate Meeting. His research interests include medical imaging, compressive sensing, and deep learning. He was a recipient of the major funding from the National Key Research and Development Program of China, the National Natural Science Foundation of China, and the Science and Technology Support Project of Sichuan Province, China. He is also a Guest Editor of International Journal of Biomedical Imaging and Sensing and Imaging and an Associate Editor of IEEE Transactions on Medical Imaging and IEEE Transactions on Radiation and Plasma Medical Sciences.
\end{IEEEbiography}

\begin{IEEEbiography}[{\includegraphics[width=1in,height=1.25in,clip,keepaspectratio]{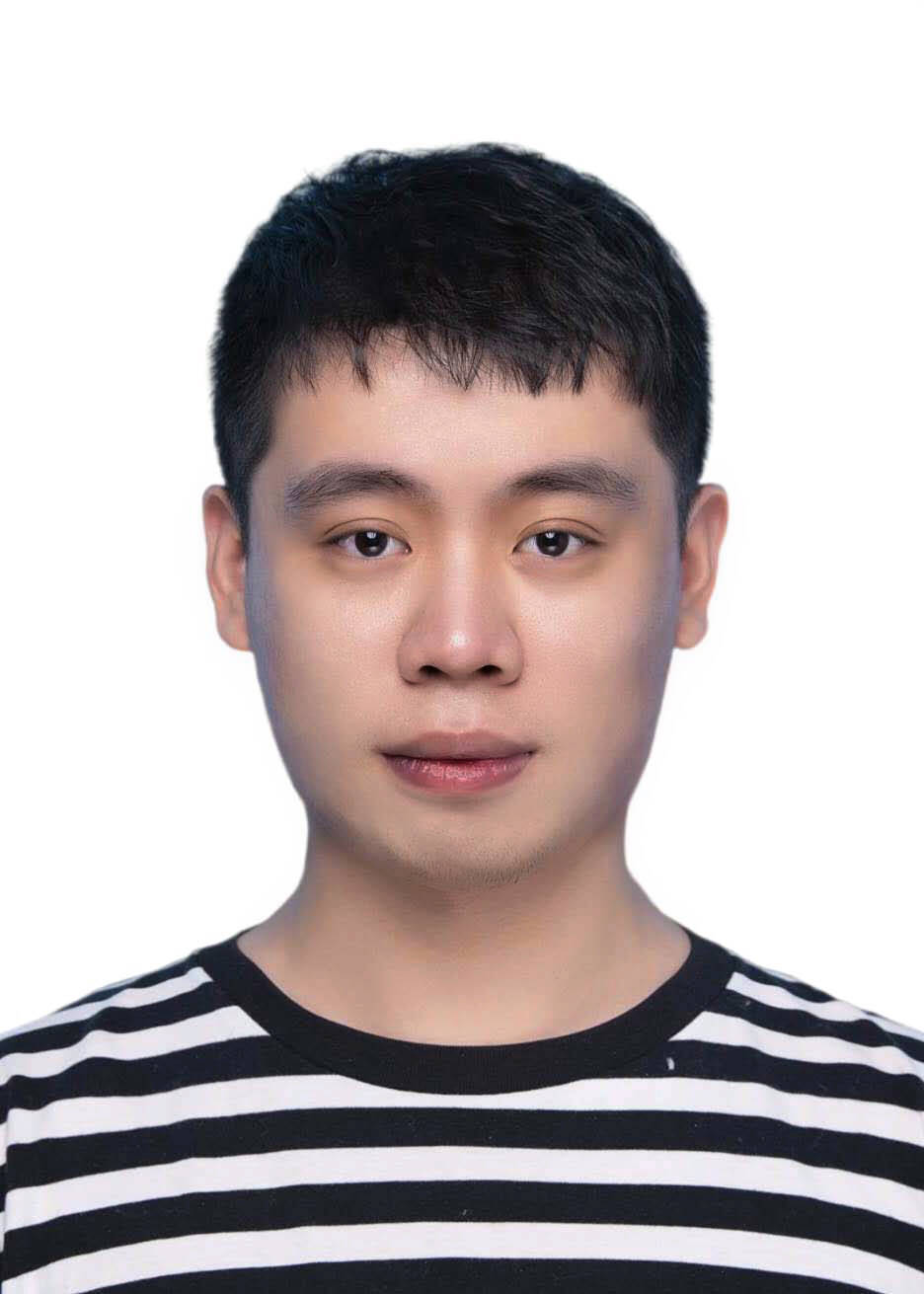}}]{Fourth D.Ziyuan Yang} is currently a postdoctoral fellow in the Department of Electronic Engineering at the Chinese University of Hong Kong. He received the M.S. degree in computer science from the School of Information Engineering, Nanchang University, Nanchang, China, in 2021, and the Ph.D. degree from the College of Computer Science, Sichuan University, China. 

He was a Research Intern at the Centre for Frontier AI Research, Agency for Science, Technology and Research (A*STAR), Singapore. In the last few years, he has published over 50 papers in leading machine learning conferences and journals, including CVPR, AAAI, IJCV, IEEE T-IFS, IEEE T-NNLS, IEEE T-SMCS, and IEEE T-AI. He was a reviewer for leading journals or conferences, e.g. IEEE T-PAMI, IEEE T-TIP, IEEE T-IFS, IEEE T-MI, CVPR, and ICCV.
\end{IEEEbiography}
\end{document}